\documentclass[manuscript]{aastex}

\def \msun   {\hbox{M$_\odot$}}
\def \rsun     {\hbox{R$_\odot$}}

\shorttitle{Stellar Diameters in the BPMG}
\shortauthors{Simon \& Schaefer}

\begin{document}

\title{Measured Diameters of Two F-stars in the Beta Pic Moving Group}

\author{M. Simon\altaffilmark{1} and G.H. Schaefer\altaffilmark{2}}

\altaffiltext{1}{Department of Physics and Astronomy, Stony Brook University,
    Stony Brook, NY 11794 michal.simon@stonybrook.edu}

\altaffiltext{2}{The CHARA Array of Georgia State University, Mount Wilson
 Observatory, Mount Wilson, CA 91023, USA schaefer@chara-array.org}

\begin{abstract}

We report angular diameters of HIP 560 and 21547, two F spectral type 
pre-main sequence members of the $\beta$~Pic Moving Group.  We used
the East-West 314-m long baseline of the CHARA Array.  The  measured 
limb-darkened angular diameters of HIP 560 and 21547 are 
$0.492\pm0.032$ ~and $0.518\pm0.009$ ~mas, respectively.   
The corresponding stellar radii are 2.1 and 1.6 \rsun ~for HIP 560
and HIP 21547 respectively. These values indicate that the stars are 
truly young.
Analyses using the evolutionary tracks calculated by Siess, Dufour, and
Forestini and the tracks of the Yonsei-Yale group yield  
consistent results. Analyzing the measurements on an angular 
diameter {\it vs} color
diagram we find that the ages of the two stars are indistinguishable; 
their average value is  $13\pm2$  MY.   The masses  of HIP 560
and 21547 are  $1.65\pm0.02$ and $1.75\pm0.05$~\msun, 
respectively.   However, analysis of the stellar parameters on 
a Hertzsprung-Russell Diagram yields ages at least 5 MY older.  
Both stars are rapid rotators.  The discrepancy between the two types 
of analyses has a natural explanation in gravitational darkening.
Stellar oblateness, however,  does not affect our measurements of
 angular diameters.

\end{abstract}


\section{Introduction}

The age and mass of a pre-main sequence (PMS) star are usually estimated
from its location in the Hertzsprung-Russell diagram (HRD) relative to
theoretical calculations of stellar evolution.   Unfortunately,
differences among the theoretical calculations can produce mass and
age estimates discrepant by factors of 2 to 3 (Hillenbrand and White,
2004; Simon, 2006).  The consequences are that the mass-spectrum of stars
produced in a star-forming region, the region's star-forming history,
and the chronology of planet formation are imprecisely known. 
The latter point has taken on a pressing importance  as astronomers are 
poised to detect exoplanets associated with young stars.

Measurement of the diameters of young stars as they contract
to the main sequence can provide their ages
and a new independent test of the theoretical PMS evolutionary
tracks (Simon, 2006).  Once a reliable set of evolutionary tracks
is established, measurement of the diameters of PMS stars will yield
measurements of their absolute ages.   
Remarkably, members of the $\beta$ Pic Moving
Group (BPMG) are young enough (10-20 MY) and near enough ($< 50$ pc;
e.g. Torres et al., 2006) that some are resolvable with the optical/IR 
interferometric array of the Center for High Angular Resolution 
Astronomy (CHARA). The brighter members of  the BPMG have 
parallaxes measured by  {\it HIPPARCOS}.  The ages of 
individual stars in the BPMG are determined by HRD fitting and the
strength  of the Li-line. An age  for the group as a whole may be estimated
dynamically by tracing the group's motion back to a common origin.  
The age estimates range from $11.2\pm 0.3$ to $22\pm 12$ MY (see 
Table 2 in Fernadez et al. 2008; Yee and Jensen, 2010).  Whether 
the spread represents  uncertainties of the dating methods
or a range in individual formation times is unknown.  Astronomers already have
evidence of planet formation associated with the members of the BPMG. 
Lagrange et al (2010) imaged the exoplanet of $\beta$ Pic, an
A-star.   Debris disks are understood as indicators of planet
formation (Quillen et al. 2007);   $\beta$ Pic and AU Mic, a BPMG 
member of M1 spectral type, are known to have debris disks  
(Smith and Terrile 1984; Lagrange et al 2010; Liu 2004).

We measured the angular diameters of two stars in the BPMG , HIP
560 and 21547, using the CHARA Array.  \S 2
describes our observations and \S 3 presents our analysis and discussion of the 
diameter measurements.  We summarize our results in \S 4.

\section{ Interferometric Measurements of  HIP 560 and HIP 21547} 

We observed at the CHARA Array  located on Mt. Wilson, CA, on
the nights of UT 11, 12, and 13 Sept. 2010.  Our observing assignment
was made through an allotment made available competitively 
to the astronomical community by CHARA and administered by the National Optical
Astronomical Observatory through its observing time application process.
The array consists of six 1-m  telescopes  in a Y-configuration 
on baselines from 34 to 331 m (ten Brummelaar et al. 2005).  We used the
CLASSIC beam combiner operating in the H and K$'$ bands to observe with
the two telescopes on the 314 m E1-W1 baseline.   At the CHARA Array
these bands are centered at 1.673 and 2.133 $\mu m$ wavelengths, 
with bandwidths, 0.285 and 0.349 $\mu m$ respectively.

Table 1 lists the properties of  HIP 560 and HIP 21547 we need for
this paper; the parallactic distances are van Leeuwen et al.'s (1997) revised
{\it HIPPARCOS} values.   Neither star is known to be a spectroscopic 
binary.  HIP 21547 is one component of a common proper motion binary; 
its companion, GJ 3305, is at about $~1'$  angular separation.   
We checked the near-IR and thermal-IR fluxes of the stars as given
by {\it 2MASS} and {\it WISE}\footnote{http://irsa.ipac.caltech.edu}.
Neither shows evidence of excess emision in the near-IR.  HIP 560 
is not included in the {\it WISE} Preliminary Data Release.  For
HIP 21547, the {\it WISE} magnitudes at 3.6, 12, and 22 ~$\mu$m
are consistent with the {\it 2MASS}
 K-band magnitude, $\sim 4.5$ (Table 1).  HIP 21547 is $\sim$ 0.5 mag
brighter in {\it WISE} band 2 at 4.5 $\mu$m; on a color-color diagram
this still places the star in the region of normal stars without a
debris disk.  Both stars are rapid rotators with values of {\it vsin
 i} in the range typical of F stars (Abt and Hunter 1962).

The essential observational datum of an interferometer measurement 
is the fringe contrast or visibility, $V$, of the target or calibrator.
An observation of a program star or its calibrators took about 10 minutes
each and consisted of a series of data scans through the central 
interferometer fringe and a sequence of shutters.  Each program star
observation was bracketed by observations of one or two stellar calibrators 
with angular diameters smaller than 0.3 mas 
and located within $10^\circ$ of the target (Table 1).   
An observation of HIP 560 or 21547 and their calibrators required 
about 2 hours to obtain good S/N. 

The calibrators were chosen to be unresolved by the interferometer, 
but their diameters are nonetheless finite (Table 1).  S. Ridgway 
(priv. comm.) kindly
estimated the limb-darkened diameters, $\phi_{LD}$, of the calibrators,
using Kervella et al's (2004) angular diameter fits for main sequence
stars as a function of their V-K color, K-band magnitude, the parallactic
distance of the star, and an estimate of its extinction (always small, $A_V \le
0.07$).   We scaled the observed calibrator visibilities to those of a 
point-like star.  We then calibrated the target visibilities by ratioing 
them to the scaled calibrator visibilities.  Table 2 lists the
wavelength, MJD time of observation, projected baseline in meters,
 its position angle on the sky, measured eastward from north,
the calibrated visibilities $V_{cal}$ of the targets, and their 
uncertainties $\sigma_{Vcal}$.
The $\sigma_{Vcal}$ include the uncertainties of the measured visibilities
of target and calibrators and an assumed $\pm 10\%$ uncertainty in the
calculated diameters of the calibrators.  

 Figures 1 and 2 show the calibrated visibilities at H and K {\it vs} 
spatial frequency (projected baseline/wavelength).  Fig. 1 for HIP 560
indicates a systematic difference in angular diameters measured at
H and K.  We do not think this can be attributed to a property of the
star because HIP 21547, also an F spectral type star, does not show
the effect.  The difference is probably measurement error.  At angular 
diameters as small as those of HIP 560 and HIP 21547 the 
curves for a uniform disk and limb-darkened disk are 
nearly  indistinguishable\footnote{This would not be true for
stars more strongly limb-darkened than F-stars (van Belle et
al. 2001).}.  Nonetheless, we did analyze the visibilities with respect to 
limb-darkened models following Hanbury-Brown et al.'s (1974)
analysis.  Their expression for the visibility of a limb-darkened
star can be written

$$ V_{LD}(x)  = [{{1-u_\lambda}\over 2}  + {{u_\lambda}\over 3}]^{-1}  
[ {1\over 2}(1-u_\lambda)V_{UD}  +{ {u_\lambda}\over {x^2}}({{ sin x}\over x} - cos x)]     $$

\parindent=0.0in

where  $V_{UD}$ is the visibility of a uniform stellar disk, 
$ V_{UD} = 2 {J_1(x) \over x}, $
in which $J_1$ is the Bessel function of order 1, and 
$x=  {{\pi \phi B} \over \lambda}   $
with $\phi$ the star's angular diameter, B the projected 
baseline during the scan, and $\lambda$ the wavelength of observation. 
We used values of the limb darkening parameter, $u_\lambda$, 
derived by Claret et al (1995) appropriate to stars that have effective temperatures
of 1.5 to 1.7 \msun ~ stars 10 to 20 MY old, $u_\lambda$ = 0.24 and
0.20,  at H and K respectively.  We included a $\pm 10\%$ uncertainty in
the values of $u_\lambda$ in the calculation of the uncertainties of
the measured stellar diameters.

\parindent=0.5in

We fit each calibrated visibility to a limb-darkened diameter, averaged
the individual values,  and present the results, 
$\overline{\phi_{Diam} (mas)}$, for 
HIP 560 and   21547 in Table 3.  The uncertainties are standard deviations 
of the mean.  The uncertainties are
dominated by the scatter of the individual visibility measurements. 
Table 3 also lists $\overline{\Phi_{Diam}}$  the ``absolute'' angular diameter,
the value of $\phi$ scaled to a common distance  of 10 pc and the
corresponding stellar radii.   The 
uncertainties in $\overline{\Phi_{Diam}}$ include the uncertainties in 
the HIPPARCOS  distances  (van Leeuven et al. 2007).

\section{Analysis and Discussion}

\subsection{Analysis in the $\Phi~vs~(V-K)$  and the HR Diagrams}

Figure 3 shows $\overline{\Phi_{Diam}}$  compared with diameters predicted
by PMS evolution models calculated by Siess et al (2000)  (SDF) and
the Yonsei-Yale (Y2) models calculated by Yi et al (2003).  The Y2
calculations\footnote{www.astro.yale.edu/demarque/yystar.html} provide
the luminosity, $L$, and effective temperature, $T_{eff}$ at 5 MY intervals
during the contraction; we calculated the corresponding radii through 
the defining relation $L=4\pi R^2 \sigma T_{eff}^4$ where $\sigma$ is
the Stefan-Boltzmann constant. The SDF website provides the photospheric 
radii directly\footnote{www-astro.ulb.ac.be/$\sim$siess/} and connects $T_{eff}$
to magnitudes using Kenyon and Hartmann's (1995) Table A5.  
 To present the Y2 results with $(V-K)$ as the independent variable, 
 we interpolated their  $T_{eff}$ to the Kenyon and 
Hartmann scale.  The diameter measurements indicate, for both stars,
ages in the range 10-15 MY  and masses 1.6 to 1.8 \msun.   The 
left-hand portion of Table 3 lists more precise values  
obtained by placing the observed diameters on a finer mesh in 
the $\Phi ~vs~ (V-K)$ color diagram (henceforth the $\Phi CD$);  the values 
lie in the columns designated $\Phi_{SDF}$ and $\Phi_{Y2}$. The age  
difference between HIP 560 and 21547 is not statistically 
significant and the SDF and Y2 tracks yield values that are in  
good agreement.  The average age of the two stars is $13\pm2$~MY.  
The masses determined using the SDF and Y2 tracks are also consistent.  
The average mass of HIP 560 is $1.65\pm0.02$ ~\msun ~and that of 
HIP 21547, $1.75\pm0.05$~\msun.

If we did not have angular diameter measurement of HIP 560 and 21547,
the only way to estimate their age and mass would be by their positions
on an HRD relative theoretical isochrones. 
Figure 4 shows such an analysis, here on an HRD
plotted as $M_K ~vs~ (V-K)$.  It is seen that both the SDF and Y2
tracks indicate ages at least 5 MY older, and masses $\sim 0.2$ \msun
~smaller than those indicated by the $\Phi CD$  (the columns 
designated HRD$_{SDF}$ and HRD$_{Y2}$ in Table 4).  

The observational inputs,  angular diameters, distances, and
photometry, and hence  the stellar radii, luminosities, and
effective temperatures give essentially the same ages and masses whether
the SDF or Y2 models are used.  The results are however 
consistently discrepant when interpreted on the
$\Phi CD$  or the HRD.   This suggests that the source of the
discrepancy lies in the application
of  theoretical calculations of non-rotating stars to stars that 
rotate with high angular velocities. 

\subsection{The Effects of Stellar Rotation}
 
The rotation of stars less massive than $\sim 2$ \msun ~slows over 
their lifetimes
because  stellar winds powered by the convective outer layers
carry away their angular momentum.  The convective zone 
in main-sequence  F spectral type stars disappears toward the earlier
F spectral type sub-classes and energy transport becomes entirely 
radiative. This
explains the rapid decrease of $vsin~i$ from F0V to
F9V (e.g. Abt and Hunter 1962; Kraft 1967).   The observation and
theoretical analysis of stellar rotation have a rich history
(e.g. Tassoul 2000)  and its effects are  revealed beautifully now that
stars can be imaged interferometrically (e.g. Peterson et al.  2006
a,b;  Monnier et al. 2007;  Zhao et al.  2009; Che et al. 2011).   

The effective surface gravity, $g_{eff}$ in a rotating star 
decreases from the pole to the
equator.  This produces oblateness and  a brightness variation with
latitude known as gravity darkening.  The oblateness of a rotating star in 
radiative equilibrium is given by

$$  o = {{R_e -R_p} \over R_e} = 0.77 {{\omega^2} \over {2\pi G
    \rho_m}}   $$

\parindent=0.0in

where $R_e$~and $R_p$ are the equatorial and polar radii, $\omega$ is the
angular velocity, and $\rho_m$ is the mean density (von Zeipel 1924; 
Chandrasekhar 1933).  Inserting numerical values,

$$ o = 0.0261 ({R_p\over \rsun})({\msun \over M_*})
({V_{eq}\over {100 km/s}})^2   $$

where $R_p$ is the polar radius of the star,  and $M_*$
and $V_{eq}$ are its mass and equatorial velocity.   
Sackmann (1970) showed that the decrease of $R_p$ from its
non-rotating value is only a few percent even when the star is rotating
nearly at break-up.   At break-up, $R_{eq} = 3/2 R_p, o = 0.33$ and
$V_{eq,bk}^2 = 2GM_*/3R_p$ ({\it eg} Collins, 1963). 
It is safe to apply these results to HIP 560 and 21547 because 
Demarque and Roeder (1967) showed that the outer convective
layers disappear at F early spectral type.

\parindent=0.5in

To make numerical estimates of the expected oblateness for 
HIP 560 and 21547, we use values from SDF for the radius and mass
of an early F spectral type star at age 13MY, $R_* = 1.7 \rsun$,
$M_*=1.7 \msun$.  We use these values because, in principle, our
measured values could be affected by oblateness.  We also use
the $vsin~i$ values, 171 and 95
km/s, measured for HIP 560 and 21547, respectively (Table 1).  
Since the inclinations are not known
we cannot calculate the equatorial velocities.  We can, however, estimate
them by calculating the expectation values  $\langle vsin~i \rangle$.
The upper bound on the inclination is $i=\pi/2$.  If the lower bound 
on $i$~were 0, $\langle vsin~i \rangle  = V_{eq}\pi/4$.  Here, however,
a lower bound, $~i_{crit}$, is set by the value at which $vsin~i_{crit}$
 ~ would imply rotation at the breakup velocity.  For the adopted 
radius and mass, $V_{eq,bk}=357$ km/s.  Hence,  $sin ~i_{crit} =$ 
~(observed v sini)/357 and $i_{crit} = 8.6^\circ$ ~and $15.4^\circ$ for 
HIP 560 and 21547 respectively.  Using these values as a lower bound on 
the inclination, the expectation values of the equatorial velocities
are 202 and 118 km/s, and of the oblateness, 0.104 and 0.036, for HIP 560
and 21547 respectively.  These values could be measured only if the stars'
rotation axes were positioned for best resolution of the oblateness with
the east-west baseline of the CHARA array.  The measured oblateness
 values are therefore likely to be smaller. In the case of HIP 21547 our
observations probably would not have measured  $o~\leq 0.036$~ 
because the precision of the diameter measurement is $\sim 2\%$ (Table 3).  
For HIP 560, in the unlikely case that a 
$\sim 10\%$ oblateness accounts for the measured value of its 
angular diameter, a $10\%$ smaller value of $\Phi$ would still place it
between the 10 and 20 MY isochrones.  We conclude that stellar 
oblateness probably does not enter into the interpretation of our 
results\footnote{Also, the on-sky angle of the projected baseline was
essentially the same throughout our observations (Table 2).  Thus
baseline rotation  during our observations is not an issue.}. 

\parindent=0.5in

In a rotating star in radiative equilibrium,  the radiative flux at 
a point on the photosphere is proportional to $g_{eff}$  
(von Zeipel 1924). Since $g_{eff}$~decreases from the pole 
to the equator,  the corresponding decrease in radiative flux can be
characterized as a decrease in the local effective temperature
and is called  gravity darkening.   In the A spectral type stars 
that have been imaged (Peterson et al. 2006a,b; Monnier et al. 2007;
Zhao et al. 2009), the temperature difference is large, $\sim 2000$K. 
Thus, relative to a non-rotating star,  a rotating star seen 
pole-on ($i=0$) will  appear brighter and its photometric color 
will be hotter  than one seen equator-on ($i=\pi/2$).
Observed values for a given star depend, of course, on its inclination.
Maeder and Peytremann (1970) have calculated the photometric effects for
stars of mass 1.4, 2, and 5 \msun ~at various values of $V_{eq}$ and
$i$.  Typically the values of $M_V$ vary by a few tenths mag  and of $(B-V)$
~by a few hundreths.  Details depend on the stellar mass and
inclination; Maeder and Peytremann's calculations show
that for $i \gtrsim 54^\circ$ the rotating star appears less bright 
and redder than a non-rotating star.

The effects of gravity darkening suggest that the discrepancies 
summarized in Table 4 are attributable to rotation of HIP 560 and 21547.
If their rotational velocities and inclinations are such that
their absolute magnitudes at K are depressed by a few tenths relative
to a non-rotating star and their $(V-K)$~ are slightly redder, the
ages and masses deduced from the $\Phi CD$ ~and HRD could be in
agreement.  A detailed test of this hypothesis will be possible when
the inclinations of these stars are measured.  This can be
accomplished by either measuring their light fluctuations and
thus rotational periods or by interferometric imaging of their
photospheric emission of these stars is mapped.  The first
approach is possible now as Garcia-Alvarez
et al. (2011) demonstrated by measuring the periods and
inclinations of two other stars in the BPMG.

\subsection{Implications for the BPMG}

If we accept the results of the $\Phi CD$ analysis and attribute the
discrepancy of results from the HRD to stellar rotation, we conclude
that HIP 560 and 21547 are roughly coeval at an average age of
about 13 MY.  Their galactic (X,Y,Z) coordinates (Table 1) place them about 49
pc apart with each moving at speed $\sim 22$ km/s given by their
(U,V,W) velocity  components.      It is too
soon to attribute the results for this pair to all the stars in the BPMG
but our results suggest that the members were born together
about 13 MY ago, and thus are at the younger end of the age span described
in \S 2. 

\section{Summary and Future Directions}

\parindent=0.0in

1)  The measured angular diameters of HIP 560 and HIP 21547 in the H
and  K bands are $0.492\pm0.032$~and $0.518\pm 0.009$ mas,
respectively.  Scaled to a common distance 10pc, their angular
diameters are $\Phi = 1.94\pm 0.13$~ and $1.52\pm 0.03$~mas.

2) Analyzing these results in $\Phi ~vs~(V-K)$ diagram with SDF and Y2
 isochrones calculated yields ages of the two stars that are not
 different at a statistically significant level.  Their average age is 
$13\pm2$ MY.  The masses determined according to the two
sets of isochrones also do not differ at a statistically different
level. The average values are $1.65\pm 0.02$ \msun~ for HIP 560
and $1.75\pm 0.05$ \msun ~for HIP 21547. 

3) Analyzing the stellar parameters with the SDF and Y2 isochrones
in a conventional  HR diagram yields ages at least 5 MY older and masses
$\sim 0.2$ \msun ~smaller.

4) HIP 560 and 21547 are rapid rotators (Table 1).  The
discrepancy in ages and masses can be accounted for by gravitational 
darkening of rotating stars in radiative equilibrium.  A detailed 
test of this hypothesis will be possible when their their inclinations 
are determined either by measuring their rotational periods or by
interometric images of their photospheres.

5) Taken together, our results suggest that HIP 560 and 21547
formed coevally about 13 MY  ago.

\parindent=0.5in

The most pressing task is to determine whether the 13 MY age
applies to the BPMG as a whole.  Although most BPMG are
in the southern hemisphere,  at least 3 other F and G spectral 
type stars are bright and near enough for measurement with the
CHARA array.  Stars of later spectral type are too faint for 
observation at the present time unless, by good luck, they
are close to the sun.  Many more of the BPMG members 
will be observable when the planned sensitivity 
improvements at the CHARA array are realized.

\vskip 1.0cm

We thank the referee for a helpful and unusually warm report.
We are grateful to Steve Ridgway for advice and help with 
the observations.  We thank Deane Peterson for reminding 
us  that F-stars can rotate rapidly, for advice about 
stellar rotation, and for pointing out the Garcia-Alvarez paper.  
We thank the CHARA staff at Mt. Wilson for thorough support. 
The CHARA Array is operated by Georgia State University's Center 
for High Angular Resolution Astronomy on Mount Wilson, California. 
Access to the CHARA Array, which is operated with funding from 
the National Science Foundation and Georgia State University, 
was obtained through a competitive TAC process administered 
by the National Optical Astronomy Observatory.  Our work was 
supported in part by NSF Grant AST-09-08406.  We used data 
products from the Two Micron All Sky Survey, which is a 
joint project of the University of Massachusetts and the 
Infrared Processing and 
Analysis Center/California Institute of Technology, funded by the 
National Aeronautics and Space Administration and the National Science 
Foundation.  Our research has also used of the SIMBAD database operated 
at CDS, Strasbourg, France.

\clearpage

\centerline{\bf References}

\parindent=0.0in

Abt, H.A. and Hunter, J.H., Jr., 1962, \apj, 136, 381

Chandrasekhar, s., 1933, MNRAS, 93, 390

Che, X. et al. 2011 \apj, 732, 68

Claret, A., Diaz-Cordoves, J., 
and Gimenez, A. 1995, \aaps, 114, 247

Collins, G.W. 1963, \apj, 138, 1134

Demarque, P. and Roeder, R. 1967 \apj, 147, 1188

Fernandez, D., Figueras, F., \& Torra, J. 2008, \aap, 480, 735

Garcia-Alvarez, D. et al. , 2011,  arXiv:1107.56882v2 

Hanbury-Brown, R., Davis, J., Lake, R.J.W., and Thompson, R.J. 
1974, MNRAS, 167, 475

Hillenbrand, L. and White, R. 2004, \apj 604, 741

Kenyon, S.~J., \& Hartmann, L. 1995, \apjs, 101, 117 

Kervella, P. et al. 2004, A\&A, 426, 297

Kraft, R. 1967, \apj, 150, 551

Lagrange, A.-M. et al. 2010, Science, 329, 57

Liu, M. 2004, Science, 305, 1442

Maeder, A. and Peytremann,E 1970, A\&A, 7, 120

Monnier, J.D. et al. 2007, Science, 317, 342

Peterson, D.M. 2006a,  Nature, 440, 896

Peterson, D.M. 2006b \apj, 636, 1087

Quillen, A.C., Morbidelli, A. and Moore, A. 2007 MNRAS, 380, 1642

Sackmann, I.-J. 1970 A\&A, 8, 76

Siess, L., Dufour, E. \& Forestini, M.  2000, \aap, 358, 593 (SDF)

Simon, M. 2006 in {\it The Power of Optical/IR 
Interferometry}, (Springer: Berlin),  A. Richichi, F. Delplancke, F. Paresce, 
\& A. Chelli, eds, p. 227

Smith, B.A. and Terrile, R.J. 1984, 226, 1421

Tassoul, J.-L. 2000 {\it Stellar Rotation} 
(Cambridge University Press: Cambridge)

ten Brummelaar, T. A., et al. 2005, \apj, 628, 453

Torres, C. A. O., Quast, G. R., 
  da Silva, L., de La Reza, R., Melo, C. H. F., Sterzik, M. 2006,
  \aap, 460, 695

van Belle, G. et al. 2001, ApJ, 559, 1155

van Leeuwen, F. et al.  1997, A\&A, 323, L61

von Zeipel, H. 1924, MNRAS, 84, 684

Yee,  J.C. and Jensen, E.L.N. 2010, \apj, 711, 303

Yi,S., Kim, Y.-C., and Demarque, P. 2003 \apjs, 144, 259

Zhao, M. et al. 2009  \apj, 701, 209

\clearpage

\begin{deluxetable}{lcc}
\tabletypesize{\scriptsize}
\tablecaption{Properties of HIP 560 and HIP 21547 \label{tbl-1}}
\tablewidth{0pt}
\tablehead{
\colhead{~} & \colhead{HIP 560}  & \colhead{HIP 21547} 
}
\startdata
Alternate Name  & HD 203          & HD 29391\\
Spectral Type     & F3V                 & F0V     \\
V(mag)               & 6.19                & 5.22    \\
K(mag)               & 5.24                & 4.54    \\ 
vsin$i$ (km/s)    & 170.7              & 95.0   \\
Distance (pc)     &$39.4\pm0.4$ &$29.4\pm0.3$\\
X,Y,Z (pc)          & 4.5, 5.8, -38.4 &-24.3, -8.2, -15.2\\
U,V,W (km/s)    & -10.4, -14.5, -13.3& -14.0, -16.2, -10.1\\ 
                &                    &                   \\
Calibrators:    &                    &                   \\
                 & HD 268             &  HD 26794         \\
$\phi_{LD}~(mas)$   &  0.27              &  0.25              \\  
                & HD 223884          &                    \\
$\phi_{LD}~(mas)$   &  0.26              &                    \\
\multicolumn{3}{c}{Data for $\beta$ Pic targets from Torres et al. (2006)}\\
\enddata

\end{deluxetable}

\begin{deluxetable}{lcccccc}
\tabletypesize{\scriptsize}
\tablecaption{Calibrated Visibilities \label{tbl-2}}
\tablewidth{0pt}
\tablehead{
\colhead{Object} & $\lambda$ & \colhead{MJD}  & \colhead{$B$ (m)} & \colhead{PA ($^\circ$)}
& \colhead{$V_{cal}$} & \colhead{$\sigma_{Vcal}$}
}
\startdata
HIP 560 & K    & 55450.286  & 308.62 & 69.69  & 0.828 & 0.018  \\
HIP 560 & K    & 55450.300  & 311.91 & 71.87  & 0.783 & 0.016  \\
HIP 560 & K    & 55450.314  & 313.44 & 73.94  & 0.751 & 0.015  \\
HIP 560 & K    & 55450.332  & 312.66 & 76.38  & 0.866 & 0.017  \\
HIP 560 & K    & 55450.347  & 309.38 & 78.47  & 0.863 & 0.016  \\
HIP 560 & K    & 55452.314  & 313.51 & 74.74  & 0.839 & 0.015  \\
HIP 560 & K    & 55452.324  & 312.91 & 76.10  & 0.814 & 0.015  \\
HIP 560 & K    & 55452.334  & 311.39 & 77.39  & 0.856 & 0.015  \\
HIP 560 & H    & 55451.328  & 312.81 & 76.22  & 0.853 & 0.028  \\
HIP 560 & H    & 55451.342  & 310.18 & 78.08  & 0.897 & 0.025  \\
HIP 560 & H    & 55451.355  & 305.80 & 79.82  & 0.893 & 0.025  \\
HIP 560 & H    & 55451.370  & 298.79 & 81.72  & 0.866 & 0.026  \\
\hline
HIP 21547 & K  & 55450.515  & 313.26 & 75.68  & 0.849 & 0.013  \\
HIP 21547 & K  & 55450.525  & 313.44 & 75.86  & 0.819 & 0.012  \\
HIP 21547 & K  & 55450.535  & 312.45 & 75.96  & 0.860 & 0.014  \\
HIP 21547 & H  & 55451.501  & 311.41 & 75.40  & 0.769 & 0.021  \\
HIP 21547 & H  & 55451.515  & 313.41 & 75.73  & 0.776 & 0.022  \\
HIP 21547 & H  & 55451.533  & 312.20 & 75.97  & 0.766 & 0.021  \\
HIP 21547 & H  & 55451.544  & 310.15 & 76.02  & 0.775 & 0.021  \\
HIP 21547 & H  & 55452.496  & 311.11 & 75.37  & 0.773 & 0.022  \\
HIP 21547 & H  & 55452.513  & 313.43 & 75.73  & 0.719 & 0.020  \\
HIP 21547 & H  & 55452.521  & 313.31 & 75.88  & 0.791 & 0.022  \\
HIP 21547 & H  & 55452.533  & 311.69 & 75.99  & 0.729 & 0.020  \\
\enddata
\end{deluxetable}

\begin{deluxetable}{lcc}
\tabletypesize{\scriptsize}
\tablecaption{Measured Sizes \label{tbl-2}}
\tablewidth{0pt}
\tablehead{
\colhead{~} & \colhead{HIP 560}  & \colhead{HIP 21547} 
}
\startdata
$\overline{\phi (mas)}$   & $0.492\pm 0.032$      &$0.518 \pm 0.009$\\
$\overline{\Phi (mas)}$   & $1.94\pm 0.13$        &$1.52 \pm 0.03$\\ 
$ R/\rsun$                & $2.08\pm 0.14$        &$1.63\pm0.03$ \\
\enddata
\end{deluxetable}

\begin{deluxetable}{lcccc}
\tabletypesize{\scriptsize}
\tablecaption{Measured Ages and Masses \label{tbl-3}}
\tablewidth{0pt}
\tablehead{
\colhead{Star} & \colhead{$\Phi_{SDF}$}  & \colhead{$\Phi_{Y2}$} 
& \colhead{HRD$_{SDF}$} & \colhead{HRD$_{Y2}$ } 
 }
\startdata
  ~ & \multicolumn{4}{c}{Age (MY)}\\
HIP 560    & $12.0\pm3.0$ & $10\pm2$&$ 25^{+25}_{-8}$ & $18\pm2$\\
HIP 21547  &$ 15.0\pm2.0$ & $15\pm2$& $20^{+30}_{-2}$&$ 18\pm2$\\
               &            &       &         &       \\
 ~ & \multicolumn{4}{c}{Mass (\msun)}\\
HIP 560    & $1.68\pm0.03$ & $1.63\pm0.02$&$1.49\pm0.05$&$1.45\pm 0.10$\\
HIP 21547  &$ 1.75\pm0.05$ &$ 1.75\pm0.05$&$1.56\pm0.05$&$1.55\pm 0.10$\\ 
\enddata
\end{deluxetable}

\clearpage
\begin{figure}
\begin{center}
\includegraphics[scale=0.8]{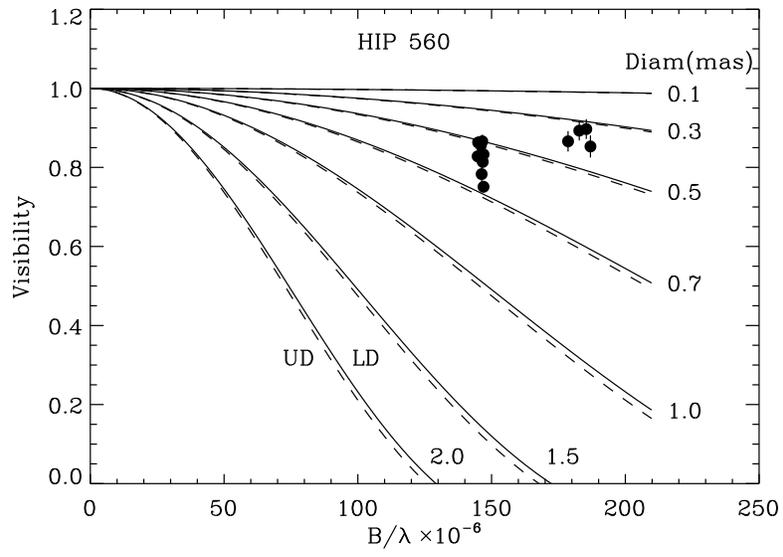}
\end{center}
\caption{Calibrated visibilities of HIP 560 at H and K {\it vs} spatial
frequency, the projected baseline divided by the wavelength. The dashed curves
are for uniform disk models (UD) with the diameters indicated and the solid
curves are limb-darkened models.}
\end{figure}

\clearpage
\begin{figure}
\begin{center}
\includegraphics[scale=0.8]{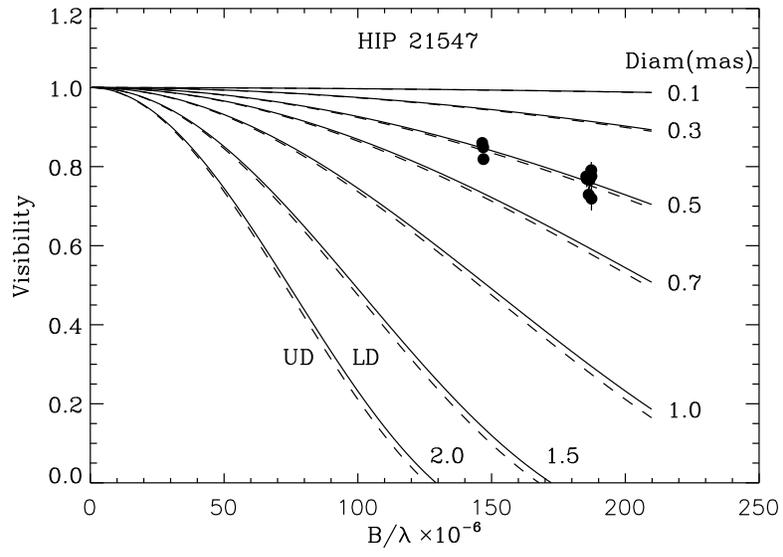}
\end{center}
\caption{Same as Fig. 1 but for HIP 21547.}
\end{figure}

\clearpage
\begin{figure}
\begin{center}
\includegraphics[scale=0.8]{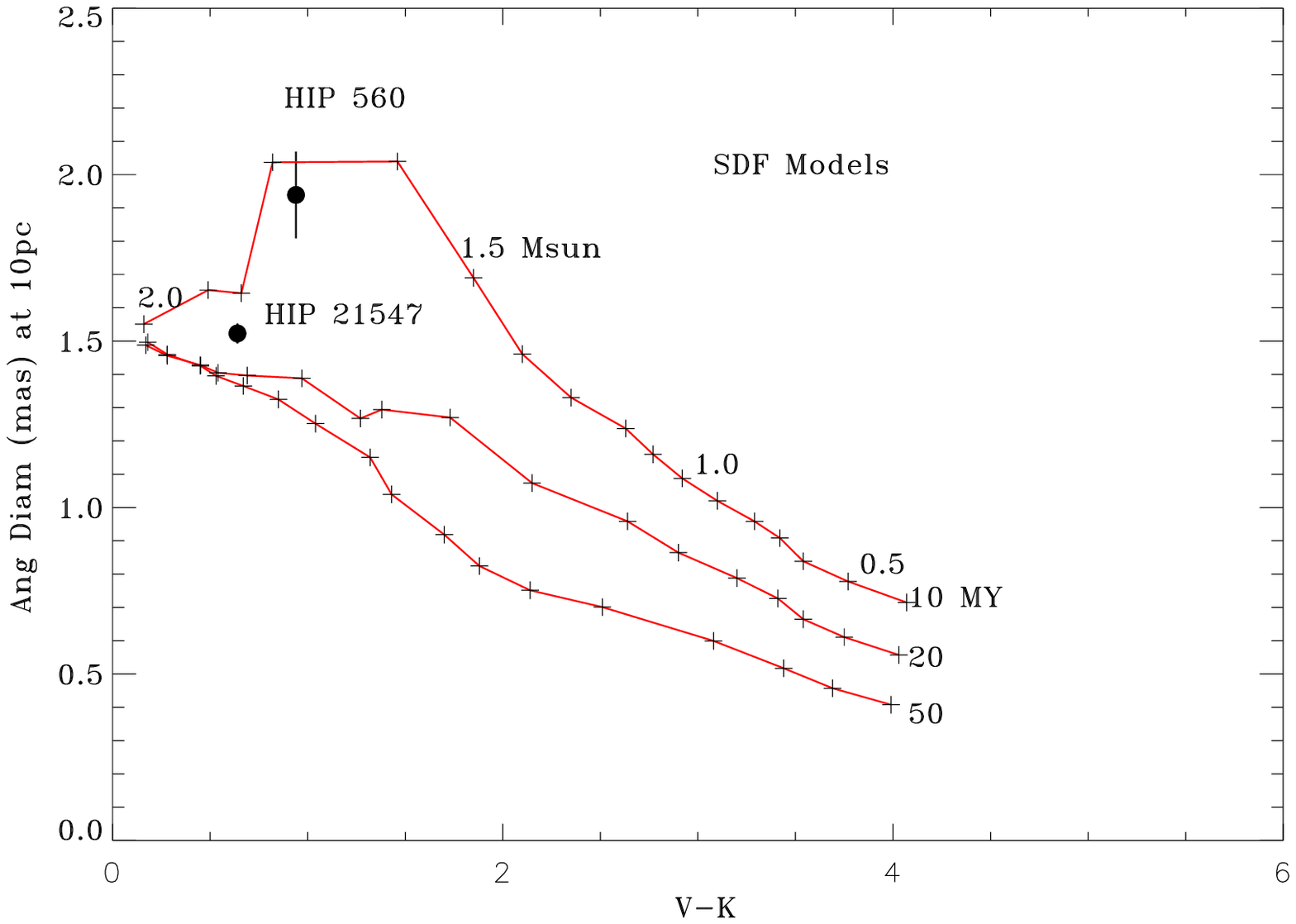}
\includegraphics[scale=0.8]{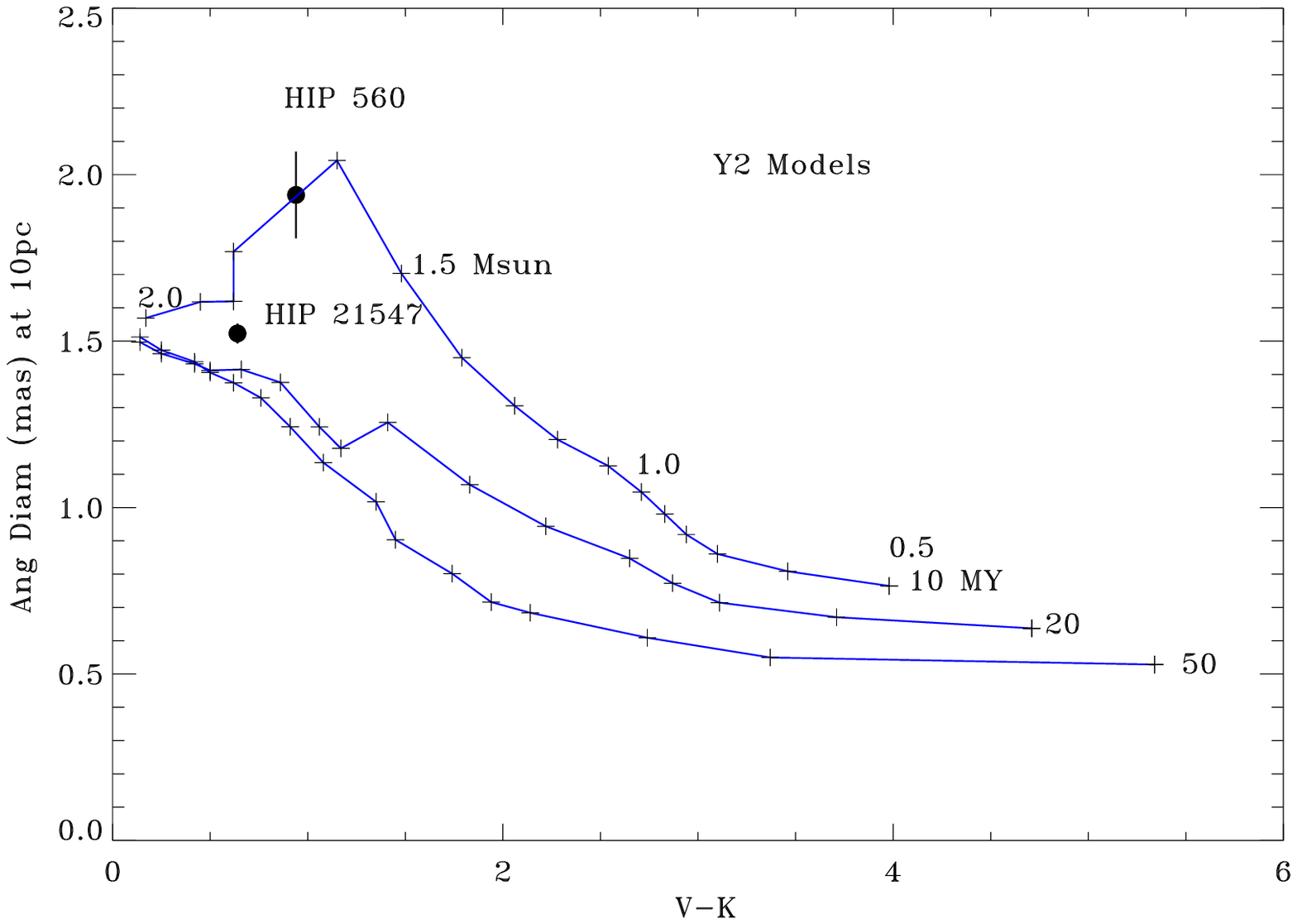}
\end{center}
\caption{HIP 560 and 21547 angular diameters scaled to 10pc distance
{\it vs}  (V-K) compared to isochrones at 10, 20, and 50 MY calculated
by SDF (top) and Y2 (bottom).  The crosses indicate masses
0.4 to 2.0 \msun~ at 0.1 \msun~intervals for the SDF isochrones and 0.5
to 2.0 \msun for the Y2 isochrones.}
\end{figure}

\clearpage
\begin{figure}
\epsscale{.80}
\plotone{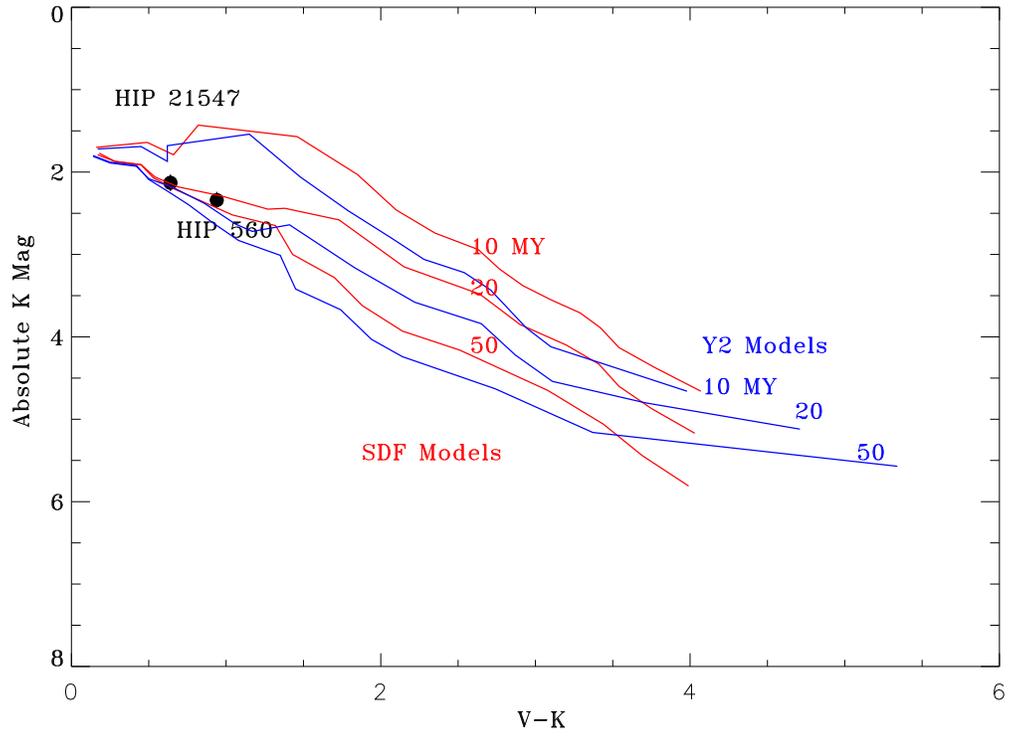}
\caption{HIP 560 and 21547 on a conventional Hetrzsprung-Russell diagram
here presented as $M_K$ {\it vs}  (V-K).  The isochrones are calculated
by SDS and Y2 as in Fig. 3.}
\end{figure}

\end{document}